\documentclass[aps,prb,10pt,twocolumn,superscriptaddress]{revtex4-2}
\usepackage{color}
\usepackage{graphicx}
\usepackage{amsmath,amsfonts,amssymb}
\usepackage{microtype}
\usepackage{float}
\usepackage{subfigure}
\usepackage{bm}
\usepackage[colorlinks=true, letterpaper=true, pdfstartview=FitV, linkcolor=red, citecolor=blue, urlcolor=blue]{hyperref}

\begin{document}
	\title{Second Chern Number Protected Gapless Vortex Modes in Superconducting Weyl Semimetals}
	
	\author{Zhicheng Zhang}\email{zhicheng_zhang1995@pku.edu.cn}
	\affiliation{School of Electronic Engineering, Guangxi University of Science and Technology, Liuzhou, Guangxi 545006, China}
	\affiliation{International Center for Quantum Materials and School of Physics, Peking University, Beijing 100871, China}
	\affiliation{School of Physical Sciences, University of Chinese Academy of Science, Beijing 100190, China}

	
	\author{Jiangping Hu}\email{jphu@iphy.ac.cn}
		\affiliation{School of Physical Sciences, University of Chinese Academy of Science, Beijing 100190, China}
	\affiliation{New Cornerstone Science Laboratory,
		Institute of Physics, Chinese Academy of Sciences, Beijing 100190, China}


	\begin{abstract}
		We have studied the vortex bound states of superconducting linear and quadratic Weyl semimetals with time reversal symmetry. For vortex with vorticity $n$ (vortex line along $z$ direction) and Weyl semimetals with Weyl point chirality $\chi$,  there are  $\vert n\chi \vert$ gapless vortex modes around the location of each Weyl point in superconducting Weyl semimetals. We show that the appearance of $\vert n\chi \vert$ gapless vortex modes originates from the   imbalanced pumping of flux bound states between Chern insulators obtained at $k_{z}$ and $-k_{z}$ in Weyl semimetals. Such a physical picture is verified by  numerical calculations on lattice models. Moreover, the gapless modes around each Weyl point can be characterized by the second Chern number defined in a synthetic four-dimensional space based on the effective BdG Hamiltonian. Finally, these gapless vortex modes are found to be robust against rotational symmetry-breaking terms. 
	\end{abstract} 
	
	
	\maketitle
	
	\section{\label{main:section1}Introduction}
	Majorana zero modes (MZMs) are fascinating objects which obey non-Abelian statistics, and the braiding of MZMs can be used to construct fault-tolerant quantum gates \cite{read2000paired,kitaev2001unpaired,KITAEV20032,RevModPhys.80.1083,wilczek2009majorana,Alicea_2012,franz2013majorana}. It is predicted that there are MZMs in the vortex cores of two-dimensional chiral $p$-wave superconductors \cite{read2000paired}.  Subsequently, it is found that there is effective $p_{x}+ip_{y}$ superconductivity on the interface of the $s$-wave superconductor-TI heterostructure \cite{PhysRevLett.100.096407}, and an MZM could be trapped at the vortex core in such a heterostructure \cite{PhysRevLett.114.017001}.  Since then, the vortex bound states (VBSs) of superconducting (SC) topological insulators (TIs) \cite{zhang2009topological,PhysRevLett.100.096407,PhysRevLett.105.097001,PhysRevB.84.144507,PhysRevLett.109.237009,hor2011superconductivity,zhang2011pressure,PhysRevLett.107.097001,PhysRevLett.114.017001,PhysRevLett.116.257003,PhysRevLett.125.037001,zhang2022bulkvortex,hu2023topological}, iron-based superconductors \cite{PhysRevLett.117.047001,PhysRevB.92.115119,PhysRevB.93.115129,qin2019topological,PhysRevLett.129.277001,PhysRevB.103.L140502,wang2018evidence,zhang2018observation,kong2019half,PhysRevB.101.020504,PhysRevLett.122.207001}, and SC Dirac semimetals \cite{PhysRevLett.108.140405,BJYang,PhysRevB.88.125427,PhysRevB.85.195320,liu2014stable,PhysRevX.11.021065,zhou2021bulk,PhysRevLett.130.046402,PhysRevB.105.014509,PhysRevLett.123.027003,PhysRevLett.124.257001,PhysRevLett.119.047001,PhysRevLett.122.207001,kobayashi2023crystal, PhysRevB.107.115127} have attracted a lot of attentions.
	
	Recent years have also witnessed remarkable progress in the study of the VBSs of SC Weyl semimetals \cite{PhysRevLett.107.127205,PhysRevLett.107.186806,PhysRevB.83.205101,nomani2023intrinsic,deng2022pressure,kononov2021superconductivity,PhysRevB.105.174502,tu2022superconductivity,PhysRevMaterials.5.084201,van2020two,PhysRevX.5.011029,huang2015weyl,li2017concurrence,PhysRevB.107.144512,PhysRevB.97.020501,PhysRevLett.118.207002,PhysRevLett.124.257001,zhang2021gapless,PhysRevLett.127.187002,PhysRevLett.130.156402,zhang2026majorana}. It is found that the vortex line can host gapless chiral Majorana modes in the pair density wave phase of a Weyl semimetal breaking time reversal symmetry \cite{PhysRevLett.118.207002}.   For time reversal symmetric Weyl semimetals with BCS pairing, it is predicted that if every constant-$k_{z}$ (the vortex is inserted along $z$ direction)  plane contains zero chirality of Weyl points, the VBSs will be fully gapped, and if certain constant-$k_{z}$ planes contain a net chirality of Weyl points, there will be gapless  modes \cite{PhysRevLett.127.187002}. The topological invariants have also been proposed to characterize the non-trivial fully gapped phase with vortex-end MZMs \cite{PhysRevLett.127.187002,PhysRevLett.124.257001}. 
	\begin{figure}[tbph]
		\centering
		\includegraphics[scale=0.313]{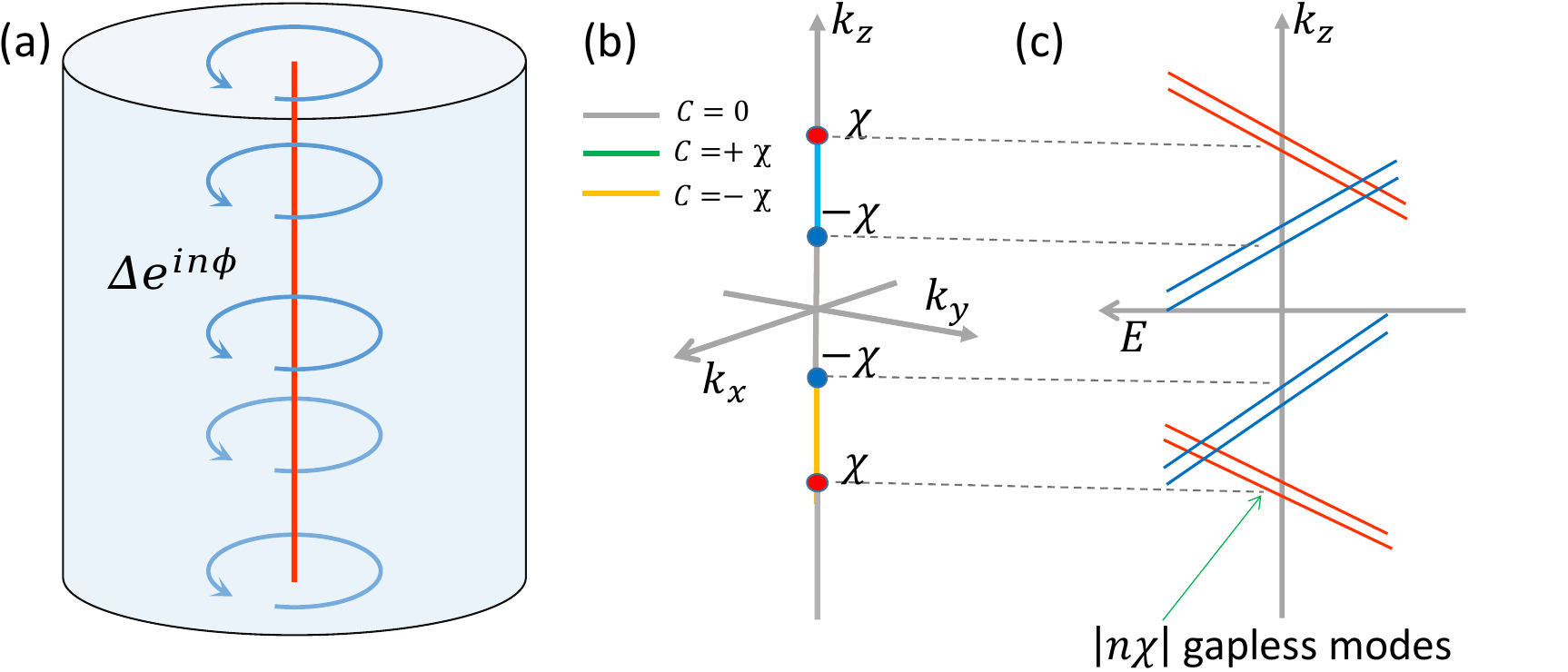}
		\caption{(a) The SC Weyl semimetal and inserted flux with vorticity $n$. (b) The four Weyl points are located on the $k_{z}$ axis, and Weyl points colored in red (blue) have the chirality $\chi$ ($-\chi$). The first Chern numbers are labelled for two-dimensional insulators at different kz. (c) There are $\vert n\chi\vert$ gapless propagating  modes around the $k_{z}$ component of each Weyl point.}
		\label{F1}
	\end{figure} 
	However, previous studies have focused on the VBSs of superconducting linear Weyl semimetals, whereas those of superconducting quadratic Weyl semimetals remain less explored. Moreover, topological invariants characterizing the gapless vortex modes in time-reversal-symmetric superconducting Weyl semimetals have yet to be proposed, and the physical origin of the appearance of these gapless  modes hasn't been thoroughly understood. Apart from the theoretical study, experimental progress about superconducting Dirac semimetals and Weyl semimetals has also been reported in the past decade \cite{hu2024observationmajoranazeromodes,kumar2024electronicstatessuperconductingtypeii,vashist2024multigapsuperconductivitynontrivialtopology,zhou2021bulk}. Some candidate materials of Weyl semimetals are found to be superconductors under high pressure \cite{li2017concurrence,deng2022pressure,PhysRevB.105.174502}, and technologies such as point contact \cite{wang2016observation} and selective ion sputtering \cite{bachmann2017inducing} have
	also been developed to induce superconductivity in Dirac semimetals and 
	Weyl semimetals. The experimental progress facilitates the study of VBSs of SC 
	Weyl semimetals and calls for further theoretical exploration.
	
	In this article, we study the VBSs of SC linear and quadratic Weyl semimetals  \cite{PhysRevLett.108.266802,huang2016new} with time reversal symmetry (TRS). As shown in Fig. \ref{F1}, we consider a flux inserted into a SC Weyl semimetal with chirality $\chi$, and $\chi=1$ ($\chi=2$) corresponds to linear (quadratic) 
	Weyl semimetal.  Fig. \ref{F1}(c) shows the schematic figure of the gapless
	modes of SC linear or quadratic Weyl semimetals. 
	We find that the numbers and propagating directions of gapless modes around each
	Weyl point can be characterized by the number $-n\cdot \chi$. There are $\vert n\chi\vert$ gapless modes around each Weyl point, and the propagating directions of the gapless modes are determined by the sign of the product $-n\chi$. The number $-n\chi$ turns out to be the second Chern number defined in a synthetic four-dimensional space based on the  low-energy effective BdG Hamiltonian. To understand the appearance of the multiple gapless  modes, we calculate the $\pi$-flux pumping of Chern insulators obtained by fixing $k_{z}$ in the models of Weyl semimetals. Based on the numerical calculations, we regard the multiple gapless  modes as the results of imbalanced $\pi$-flux pumping of Chern insulators at constant $k_{z}$ and $-k_{z}$ planes. Finally, we find that the multiple gapless  modes still exist after adding rotational symmetry-breaking terms, which shows the robustness of these gapless  modes. 
	
	\section{\label{main:section2}Model}
	In this section, we write down the $\bm{k}\cdot\bm{p}$ model of linear and quadratic Weyl semimetals. 
	First, the continuous model of the linear Weyl semimetal  with TRS, and  $C^{z}_{4}$ rotational symmetry is written under the basis $(\vert+,\frac{1}{2} \rangle, \vert-,\frac{3}{2} \rangle, \vert+,-\frac{1}{2} \rangle, \vert-,-\frac{3}{2} \rangle)^{T}$, in which each state is the eigenstate of inversion and $C^{z}_{4}$ operations labelled by the corresponding eigenvalues. 
	\begin{equation}
		\begin{aligned}
			H_{\rm{LWSM}}(\bm{k})=\left( \begin{matrix} M_{1}(\bm{k})& Ak_{+}&0 & B^{*}(\bm{k})\\  Ak_{-}& -M_{1}(\bm{k})& B^{*}(\bm{k}) &0 \\ 0 &B(\bm{k}) &M_{2}(\bm{k}) &-Ak_{-}\\  B(\bm{k}) &0 &-Ak_{+} &-M_{2}(\bm{k})  \end{matrix} \right)
		\end{aligned}.
		\label{E1}
	\end{equation}
	Here $M_{1,2}(\bm{k})=M_{0}-M_{1}k^{2}_{z}-M_{2}(k^{2}_{x}+k^{2}_{y})\pm \eta k_{z}$, and $B(\bm{k})=\beta k_{z}(k^{2}_{x}-k^{2}_{y})+2i\alpha k_{x}k_{y}k_{z}$. As shown in Fig. \ref{F1}(b), the gap-closing condition of the above Hamiltonian gives four Weyl points with chirality $\chi=1$ located on $k_{z}$ axis. The  corresponding  tight-binding lattice  model is given in the supplemental materials \cite{SM}, in which we also give the correspondence between the parameters in $\bm{k}\cdot \bm{p}$ model and the parameters in tight-binding lattice model. 
	
	The  $\bm{k}\cdot \bm{p}$  model of the quadratic Weyl semimetal is written under the basis $(\vert+,\frac{5}{2} \rangle, \vert-,\frac{1}{2} \rangle, \vert+,-\frac{5}{2} \rangle, \vert-,-\frac{1}{2} \rangle)^{T}$, in which each state is the eigenstate of inversion and $C^{z}_{6}$ operations labelled by the corresponding eigenvalues. 
	\begin{equation}
		H_{\rm{QWSM}}=\left( \begin{matrix} M_{1}(\bm{k})& Ak^{2}_{-}k_{z} &0 & B^{*}(\bm{k})\\  Ak^{2}_{+}k_{z}  & -M_{1}(\bm{k})& B^{*}(\bm{k}) &0 \\ 0 &B(\bm{k}) &M_{2}(\bm{k}) &-Ak^{2}_{+}k_{z} \\  B(\bm{k}) &0 &-Ak^{2}_{-}k_{z}  &-M_{2}(\bm{k}  \end{matrix} \right).
		\label{E2}
	\end{equation}
	\begin{figure}[htbp]
		\centering
		\includegraphics[scale=0.26]{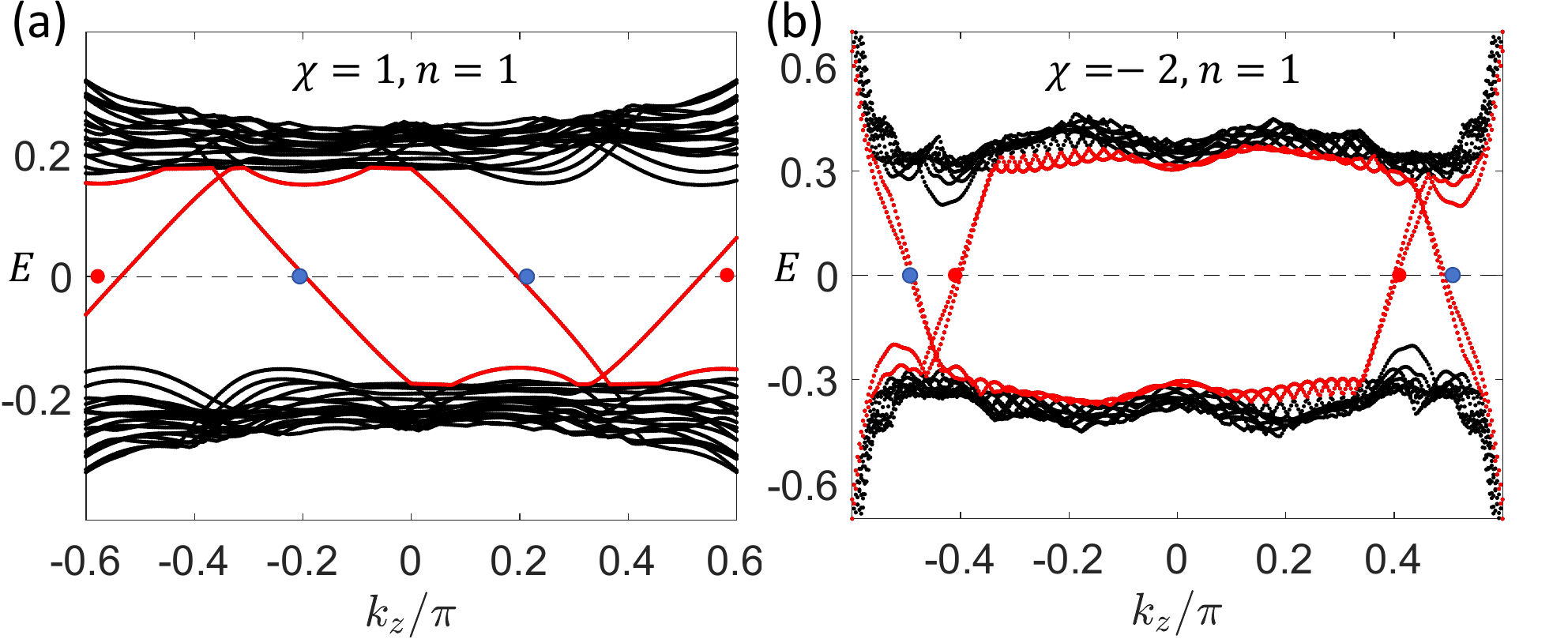}
		\caption{The VBSs of SC Weyl semimetals. The red and blue points in the two sub-figures are Weyl points with opposite chirality. (a) The VBSs ($E-k_{z}$ relation) of SC linear Weyl semimetal are calculated with $\mu=0.2$, $\eta=0.3$, and $n=1$. (b) The VBSs of SC quadratic Weyl semimetal  are calculated with $\mu=0.4$, $\eta=0.7$, and $n=1$.} 
		\label{F2}
	\end{figure} 
	In the above Hamiltonian Eq. (\ref{E2}), we have  $M_{1,2}(\bm{k})=M_{0}-M_{1}k^{2}_{z}-M_{2}(k^{2}_{x}+k^{2}_{y})\pm \eta k_{z}$, and $B(\bm{k})=\alpha k_{x}(k^{2}_{x}-3k^{2}_{y})+i\beta k_{y}(3k^{2}_{x}-k^{2}_{y})$.
	Eq. (\ref{E2}) describes a quadratic Weyl semimetal with four
	quadratic Weyl points with chirality $\chi=-2$ located on the $k_{z}$ axis. The corresponding tight-binding lattice model is given in the supplemental materials \cite{SM}.

	The bases of the $\bm{k}\cdot\bm{p}$ models of linear and quadratic Weyl semimetals can be written in a unified form $\psi(\bm{k})=(C_{A,l}(\bm{k}), C_{B,m}(\bm{k}), C_{A,-l}(\bm{k}), C_{B,-m}(\bm{k}))$, in which  $A$/$B$ label the parity, and $l$/$m$ label the angular momentum $L_{z}$. When we consider the Hamiltonian describing SC  Weyl semimetals, we choose the Nambu basis as $\Psi({\bm{k}})=(\psi(\bm{k}),\psi^{\dagger}(-\bm{k}))^{T}$. Under this basis, the BdG Hamiltonian of the SC Weyl semimetal is  $H=\sum_{\bm{k}} \Psi^{\dagger}({\bm{k}}) H_{\rm{BdG}}(\bm{k}) \Psi({\bm{k}})$, where 
	\begin{equation}
		H_{\rm{BdG}}(\bm{k})=\left( \begin{matrix} H_{0}(\bm{k})-\mu & \Delta \\  \Delta^{\dagger} & \mu-H^{*}_{0}(-\bm{k}) \end{matrix}\right).
		\label{E3}
	\end{equation}
	In the BdG Hamiltonian Eq. (\ref{E3}), $H_{0}(\bm{k})$ is the normal state Hamiltonian of linear or quadratic Weyl semimetal. The vortex line adds the phase $e^{i\theta}$ ($\theta=\arctan y/x$) to the pairing potential and preserves the translation symmetry in $z$ direction.  
	Thus, $k_{z}$ is a good quantum number, and we can calculate the VBSs for each $k_{z}$ with open boundary condition in $x$ and $y$ directions. We consider the $s$-wave singlet pairing and assume that the pairing potential at the vortex core is $\Delta=0$, and $\Delta(r)=\Delta_{0}e^{in\theta}$ on other lattice sites. For SC linear Weyl semimetal, the parameters in the lattice model are set as $\{t^{D}_{1},t^{C}_{4}, m, t^{A}_{1}, t^{A}_{3}, t^{B}_{4}, \Delta_{0}\}=\{-0.25, -0.125, 2.2, -0.5, 0.5, 0.25, 0.2 \}$, and for SC quadratic Weyl semimetal, the parameters in the lattice model are set as $\{t^{D}_{1}, t^{A}_{1}, m, t^{A}_{4}, t^{B}_{4}, t^{C}_{3}, \Delta_{0}\}=\{-2.5, -1.4, 8.4, -1.0, -0.46, 0.6, 0.4\}$.  
	
	\section{\label{main:section3}The VBSs Characterized by the Second Chern Number}
	We have performed numerical calculations of the VBSs of SC linear and quadratic Weyl semimetals, and the results are shown in Fig. \ref{F2}.   Concretely, Fig. \ref{F2}(a) shows the VBSs of SC linear Weyl semimetal ($\chi=1$) with vorticity $n=1$. Fig. \ref{F2}(b)  shows the VBSs of SC quadratic Weyl semimetal ($\chi=-2$) with vorticity $n=1$. The red and blue points in the two sub-figures label the $k_{z}$ coordinates of Weyl points with opposite chirality. We notice that the gapless modes around each Weyl point can be characterized by the product $-n\chi$. To be specific, the number of gapless  modes around the $k_{z}$ component of each Weyl point is $\vert  n\chi \vert $, and the propagating direction of gapless  modes around each Weyl point is determined by the sign of the product $-n\chi$. For example, there is $\vert n\chi\vert=1$ gapless mode around each linear Weyl point in Fig. \ref{F2}(a), while there are $\vert n\chi\vert=2$ gapless modes around each quadratic Weyl point in Fig. \ref{F2}(b). Besides, the propagating direction of gapless modes around each Weyl point in Fig. \ref{F2}(b) is opposite to that in Fig. \ref{F2}(a), which is consistent with the fact that we have $\chi=-2$ in Fig. \ref{F2}(b) and $\chi=1$ in Fig. \ref{F2}(a).  In fact, in the supplemental materials \cite{SM}, there are more numerical examples demonstrating that the dependence of propagating direction and number of gapless  modes on $n$ and $\chi$ can be characterized by the product $-n\chi$.
	
	We find that the product $-n\chi$ is nothing but the emergent second Chern number \cite{PhysRevB.82.115120,bernevig2013topological,PhysRevB.78.195424,zhang2001four} defined in a synthetic four-dimensional space that stems from the effective BdG Hamiltonian \cite{PhysRevB.86.214514,PhysRevB.86.054504}. For simplicity, we fix $\mu=0$ in the following discussions. In this case, the normal-state Fermi surface consists of isolated Weyl points. To derive the 
	effective BdG Hamiltonian, we focus on the low-energy physics around the four Weyl points. The detailed derivations of the effective BdG Hamiltonian are given in the supplementary materials \cite{SM}. If we use $q_{i}$ $(i=x,y,z)$ to denote the momentum relative to a Weyl point and assume $q_{\pm}=q_{x}\pm iq_{y}=qe^{\pm i\theta}$, the
	final form of the effective BdG Hamiltonian is
	\begin{equation}
		\begin{aligned}
			H_{\rm{BdG}}(\bm{q})&=\left( \begin{matrix}  vq_{z} & Aq^{\chi}_{-} &\Delta e^{in\phi}&0\\  Aq^{\chi}_{+} & -vq_{z} &0&\Delta e^{in\phi}\\ \Delta e^{-in\phi}&0&-vq_{z} & -Aq^{\chi}_{-} \\ 0&\Delta e^{-in\phi}& -Aq^{\chi}_{+} & vq_{z}  \end{matrix}\right).
			\label{E4}
		\end{aligned}
	\end{equation}
	For a region far away from the vortex core, we may assume it is a three-dimensional space with approximate constant $\phi$. Then combining all such regions together leads to a synthetic four-dimensional ($q_{x}$, $q_{y}$, $q_{z}$, $\phi$) space with periodic boundary condition in parameter $\phi$ \cite{PhysRevLett.119.047001,PhysRevB.82.115120}.  Therefore, we can define the second Chern number $C_{2}$ in this synthetic four-dimensional space.
	\begin{figure}[htbp]
		\centering
		\includegraphics[scale=0.42]{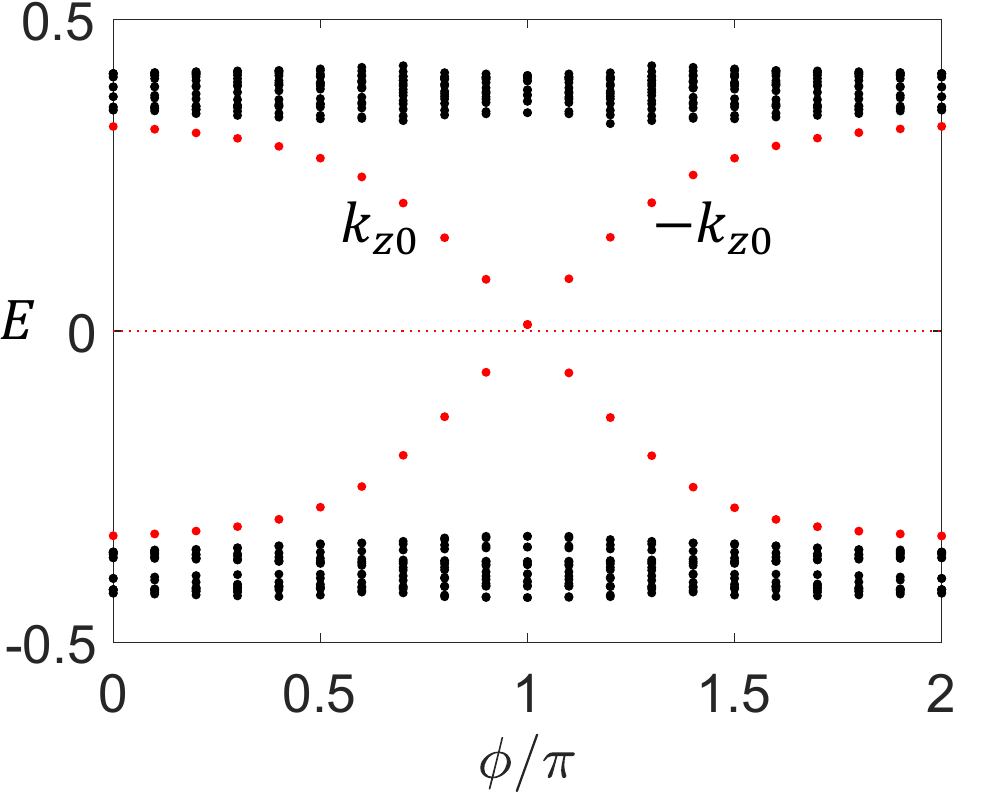}
		\caption{Understanding the appearance of the gapless  modes from the $\pi$-flux pumping of  Chern insulators in a linear Weyl semimetal. The red dots are flux bound states localized at the flux core.  This figure shows the pumping of flux bound states of two-dimensional insulators at $k_{z0}$ $(0.35\pi)$ and $-k_{z0}$ as the inserted flux increases from $\phi=0$ to $\phi=2\pi$. The results are calculated with $\mu=0$ and $\eta=0.3$.} 
		\label{F3}
	\end{figure} 
	
	To calculate $C_{2}$ of the above effective BdG Hamiltonian, we reformulate Eq. (\ref{E4}) as
	\begin{equation}
		\begin{aligned}
			H_{\rm{BdG}}&=Aq^{\chi}\cos(\chi\theta)\chi_{3}\otimes \tau_{1}+Aq^{\chi}\sin(\chi\theta)\chi_{3}\otimes \tau_{2} \\
			&+vq_{z}\chi_{3}\otimes \tau_{3} -\Delta \sin (n\phi) \chi_{2}\otimes \tau_{0}\\
			&+\Delta \cos (n\phi) \chi_{1}\otimes \tau_{0}.
			\label{E45}
		\end{aligned}
	\end{equation}
	Then we define $d_{1}=Aq^{\chi}\cos \chi\theta$, $d_{2}=Aq^{\chi}\sin  \chi\theta$, $d_{3}=vq_{z}$, $d_{4}=-\Delta \sin n\phi$, $d_{5}=\Delta \cos n\phi$. In this way, we can write the effective BdG Hamiltonian as $H_{\rm{BdG}}(\bm{q})=\sum_{i} d_{i}(\bm{q}, \phi)\cdot \Gamma_{i}$, in which $\Gamma_{i}$ are five mutually anti-commutating matrices in Eq. (\ref{E45}). The definition of the second Chern number is
	\begin{equation}
		\begin{aligned}
			C_{2}&=\frac{3}{8\pi^{2}} \int d^{3}q \int d\phi  \frac{1}{d^{5}} \epsilon^{abcde} d_{a}\partial_{q}d_{b}\partial_{\theta}d_{c}\partial_{q_{z}}d_{d}\partial_{\phi}d_{e}.
			\label{E5}
		\end{aligned}
	\end{equation}
	Here we have defined  $d=(\sum_{i}d^{2}_{i})^{1/2}$.
	If we substitute the explicit forms of the coefficients $d_{i}$ into the definition of the second Chern number, we will obtain $C_{2}=-n\cdot \chi$ after some
	algebra \cite{SM}. This is the expected topological invariant which describes the dependence of gapless  modes on the vorticity $n$ and chirality $\chi$ in SC linear and quadratic Weyl semimetals.
	
	\section{\label{main:section4}The Pumping Picture}
	Now we provide a physical picture to understand such dependence of the
	VBSs on $n$ and $\chi$. As shown in Fig. \ref{F1}(b), the Weyl semimetals we consider have four Weyl points on $k_{z}$ axis, and Chern numbers can be defined for the two-dimensional insulators obtained by fixing $k_{z}$ in the model of Weyl semimetals. Thus, the Weyl semimetals can be regarded as Chern insulators with different Chern numbers stacked along $z$ direction. The imbalanced pumping between Chern insulators at $k_{z}$ and $-k_{z}$ in Weyl semimetals can lead to the appearance of $\vert n\chi \vert$ gapless vortex modes around the location of each Weyl point in the VBSs of SC Weyl semimetals.
	Now we take the VBSs of SC linear Weyl semimetal as an example, and we consider inserting a flux into the linear Weyl semimetal along $z$ direction. Fig. \ref{F3} shows the pumping of the spectra of Chern insulators at $k_{z0}$ and $-k_{z0}$. Here $k_{z0}=0.35\pi$ is a typical value of $k_{z}$, at which the insulator has nonzero Chern number. The red dots are states which localize around the flux core, and they are dubbed \textit{flux bound states}.  Because of the existence of TRS, Chern insulators at $k_{z}$ and $-k_{z}$ have opposite Chern numbers.  As a result, we can find that the flux bound states at $k_{z0}$ cross the Fermi level downwards, while the flux bound states at $-k_{z0}$ cross the Fermi level upwards. In fact, there is also the pumping of edge state. But to highlight the pumping of flux bound states, we don't plot the edge states in this figure.
	
	Although Fig. \ref{F3} shows the $2\pi$-flux pumping, we should focus on the $\pi$-flux pumping to understand the VBSs of SC linear Weyl semimetals. Combining the $\pi$-flux pumping of the flux bound states of two Chern insulators at
	$k_{z0}$ and $-k_{z0}$, we conclude that depending on the value of chemical potential, the flux bound states at $-k_{z0}$ are pumped from occupied to unoccupied, or the flux bound states at $k_{z0}$ are pumped from unoccupied to occupied. In the BdG Hamiltonian, there are both particle and hole degrees of freedom \cite{zhang2021gapless}, so we have to consider the hole counterpart of the aforementioned pumping process. It turns out that for the $\pi$-flux pumping in the hole part, we also obtain the result that the flux bound states at $-k_{z0}$ are pumped from occupied to unoccupied, or the flux bound states at $k_{z0}$ are pumped from unoccupied to occupied.  We assume that the pumping of flux bound states won't be qualitatively changed in the weak pairing limit. Consequently, we have $N_{sc}(k_{z0})-N_{sc}(-k_{z0})=2$, where $N_{sc}(k_{z})$ is the number of negative energy vortex bound states at $k_{z}$.  Meanwhile, as shown in Fig. \ref{F2}(a), we notice that around $k_{z}=0$ we have $N_{sc}(k_{z})-N_{sc}(-k_{z})=0$, because for insulators around $k_{z}=0$ with Chern number $C=0$, there is no the pumping of flux bound states. However, if $k_{z}$ is greater than the $k_{z}$ coordinate of the right blue Weyl point in Fig. \ref{F2}(a), we will have $N_{sc}(k_{z})-N_{sc}(-k_{z})=2$ according to the aforementioned argument. So we expect that there is a positive-energy VBS crossing the Fermi level and becoming a negative-energy VBS, resulting $\vert n\chi \vert=1$ gapless vortex mode around the right blue Weyl point.  Similarly, we can argue that there is $\vert n\chi\vert$=1 gapless  mode around the $k_{z}$ component of each of the four Weyl points.

	Analogous to the pumping picture above, we can readily understand the dependence of the number of gapless  modes on the chirality $\chi$ and the vorticity $n$. For instance, for a quadratic Weyl semimetal with chirality $\chi=-2$, we can obtain $C=2$ Chern insulator at an appropriate value of $k_{z}$. 
	Similar to Fig. \ref{F3}, after inserting a flux with vorticity $n=1$, there will be two flux bound states pumped from occupied to unoccupied at $-k_{z0}$, or two flux bound states pumped from unoccupied to occupied at $k_{z0}$.
	Correspondingly, in the weak pairing limit we have $N_{sc}(k_{z0})-N_{sc}(-k_{z0})=4$, and there will be $\vert n\chi\vert=2$ gapless  modes around each Weyl point. On the other hand, if we keep $\chi=1$ (linear Weyl semimetal) and change the vorticity from $1$ to $2$, there are effectively two $n=1$ fluxes. As a result, there will also be two flux bound states pumped from occupied to unoccupied at $-k_{z0}$, or two flux bound states pumped from unoccupied to occupied at $k_{z0}$. Therefore, the number of gapless  modes around each Weyl point also doubles. To sum up, we have shown that the number of gapless  modes around each Weyl point is proportional to the product $\vert n\chi\vert$ through the pumping picture.
	\begin{figure}[htbp]
		\centering
		\includegraphics[scale=0.26]{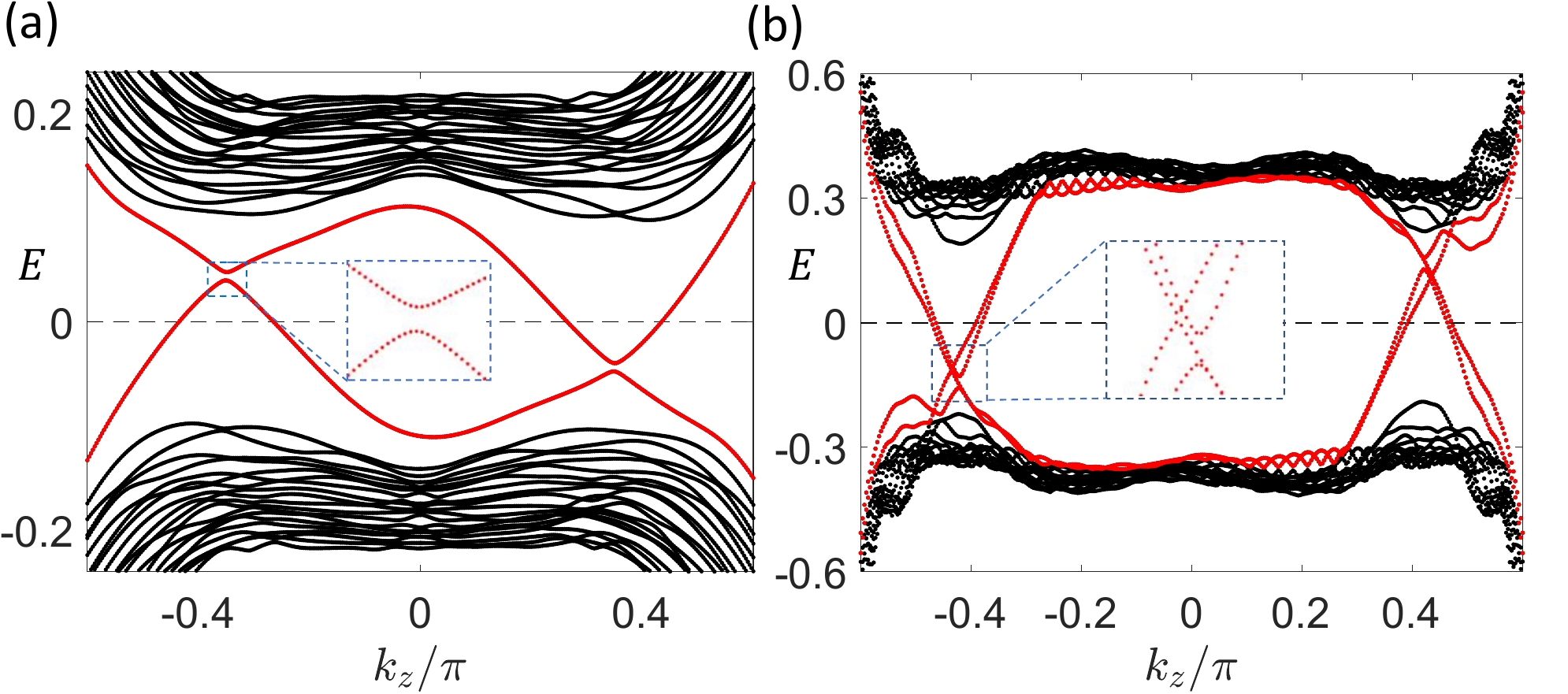}
		\caption{The robustness of multiple gapless  modes against rotational symmetry-breaking perturbations. (a) The VBSs of SC linear Weyl semimetal without uniaxial rotational symmetry. Here we set $\mu=0.2$, $\eta=0.1$, $\epsilon=0.1$, and $n=1$. (b) The VBSs of SC quadratic Weyl semimetal without uniaxial rotational symmetry. The parameters are set as $\mu=0.6$, $\eta=0.7$, $\epsilon=0.2$, and $n=1$.}
		\label{F4}
	\end{figure}

	\section{\label{main:section5}The Robustness of Gapless  Modes}
	The models of linear and quadratic Weyl semimetals above are written based on models of linear and quadratic Dirac semimetals, and they are rotational symmetric linear ($C_{4}$) and quadratic Weyl semimetals ($C_{6}$). However, it is not necessary for a Weyl semimetal to be rotational symmetric. So it is natural to ask whether the multiple gapless  modes around the Weyl points are robust against rotational symmetry-breaking perturbations. To answer this question, we add the rotational symmetry breaking term $\epsilon \sigma_{3}\otimes \tau_{2}$ into both the linear and quadratic Weyl semimetals. As a result, the four Weyl points are no longer located on the $k_{z}$ axis. Instead, they are now located on the $k_{y}-k_{z}$ plane. The corresponding VBSs of rotational symmetry-breaking SC linear and quadratic Weyl semimetals are shown in Fig. \ref{F4}(a) and Fig. \ref{F4}(b), respectively. We find from the inset in Fig. \ref{F4}(a) that there is still one propagating gapless  mode around each Weyl point, but the intersection of two counter-propagating   modes is gapped by the rotational symmetry-breaking term. Similarly,  Fig. \ref{F4}(b) shows that there are still two propagating gapless  modes around each Weyl point, while the intersections of four counter-propagating  modes are gapped by the rotational symmetry-breaking perturbation.
	
	 Although the intersection of counter-propagating  modes are gapped, they are far away from Fermi energy. Since the main properties of a superconductor are determined by the low-energy spectra near the Fermi energy, the spectra around the intersections don't play the main role in determining the physical properties. In this sense, we claim that the spectra in Fig. \ref{F4} are still gapless, and the multiple gapless propagating  modes around each Weyl point are robust against rotational symmetry-breaking perturbations.

	\section{Summary and Discussion}
	In summary, we have studied the VBSs of SC Weyl semimetals with time reversal symmetry. We show that the imbalanced pumping of flux bound states between Chern insulators obtained at $k_{z}$ and $-k_{z}$ in Weyl semimetals can lead to the appearance of $\vert n\chi \vert$ gapless modes around each Weyl point in the VBSs of SC Weyl semimetals.  Such a physical picture is verified by  numerical calculations based on lattice models. We also conclude from the numerical results that the gapless modes around each Weyl point can be characterized by the product $-n\chi$, which is the second Chern number defined in a synthetic four-dimensional space spanned by parameters $q_{x}$, $q_{y}$, $q_{z}$, and $\phi$. Finally, we show these low-energy gapless modes are robust against rotational symmetry-breaking terms. 
	
	In this article, we have also used the pumping picture to understand the origin of gapless modes in the vortex line of SC Weyl semimetals. As a matter of fact, because the pumping of flux bound state is a generic feature for topological bands, it is natural to expect that topological VBSs will appear in not only SC Weyl semimetals but also SC topological insulators and SC Dirac semimetals, which have been extensively studied in recent years \cite{PhysRevLett.124.257001,PhysRevLett.127.187002,PhysRevLett.107.097001,PhysRevLett.123.027003}. Thus, the flux-pumping picture of the normal-state topological models may provide a unified framework for understanding VBSs in corresponding superconducting topological materials. A systematic exploration of this possibility is left for future work.

	\section*{ACKNOWLEDGEMENTS}
	We thank Shengshan Qin and Professor Xiong-Jun Liu sincerely for helpful discussions. J.P. Hu  is supported by the Ministry of Science and Technology  (Grant No. 2022YFA1403900), 
	the National Natural Science Foundation of China (Grant No. NSFC-12494594),  and the New Cornerstone Investigator Program.
	
	\bibliography{main_refs}
	\bibliographystyle{apsrev4-2}
\end{document}


\renewcommand{\thesection}{S-\arabic{section}}
\setcounter{section}{0}  
\renewcommand{\theequation}{S\arabic{equation}}
\setcounter{equation}{0}  
\renewcommand{\thefigure}{S\arabic{figure}}
\setcounter{figure}{0}  
\renewcommand{\thetable}{S\Roman{table}}
\setcounter{table}{0}  

\title{Supplemental Materials: Second Chern Number Protected
Gapless Vortex Modes in Superconducting Weyl Semimetals}

\author{Zhicheng Zhang}
\email{zhicheng_zhang1995@pku.edu.cn}

\affiliation{School of Electronic Engineering,
Guangxi University of Science and Technology,
Liuzhou, Guangxi 545006, China}

\affiliation{International Center for Quantum Materials
and School of Physics, Peking University,
Beijing 100871, China}

\affiliation{School of Physical Sciences,
University of Chinese Academy of Science,
Beijing 100190, China}

\author{Jiangping Hu}\email{jphu@iphy.ac.cn}
		\affiliation{School of Physical Sciences, University of Chinese Academy of Science, Beijing 100190, China}
	\affiliation{New Cornerstone Science Laboratory,
		Institute of Physics, Chinese Academy of Sciences, Beijing 100190, China}

\maketitle

\onecolumngrid
\flushbottom
\section{The Effective BdG Hamiltonians of SC Weyl Semimetals and Emergent Second Chern Number}
\begin{figure}[htbp]
	\centering
	\includegraphics[scale=0.32]{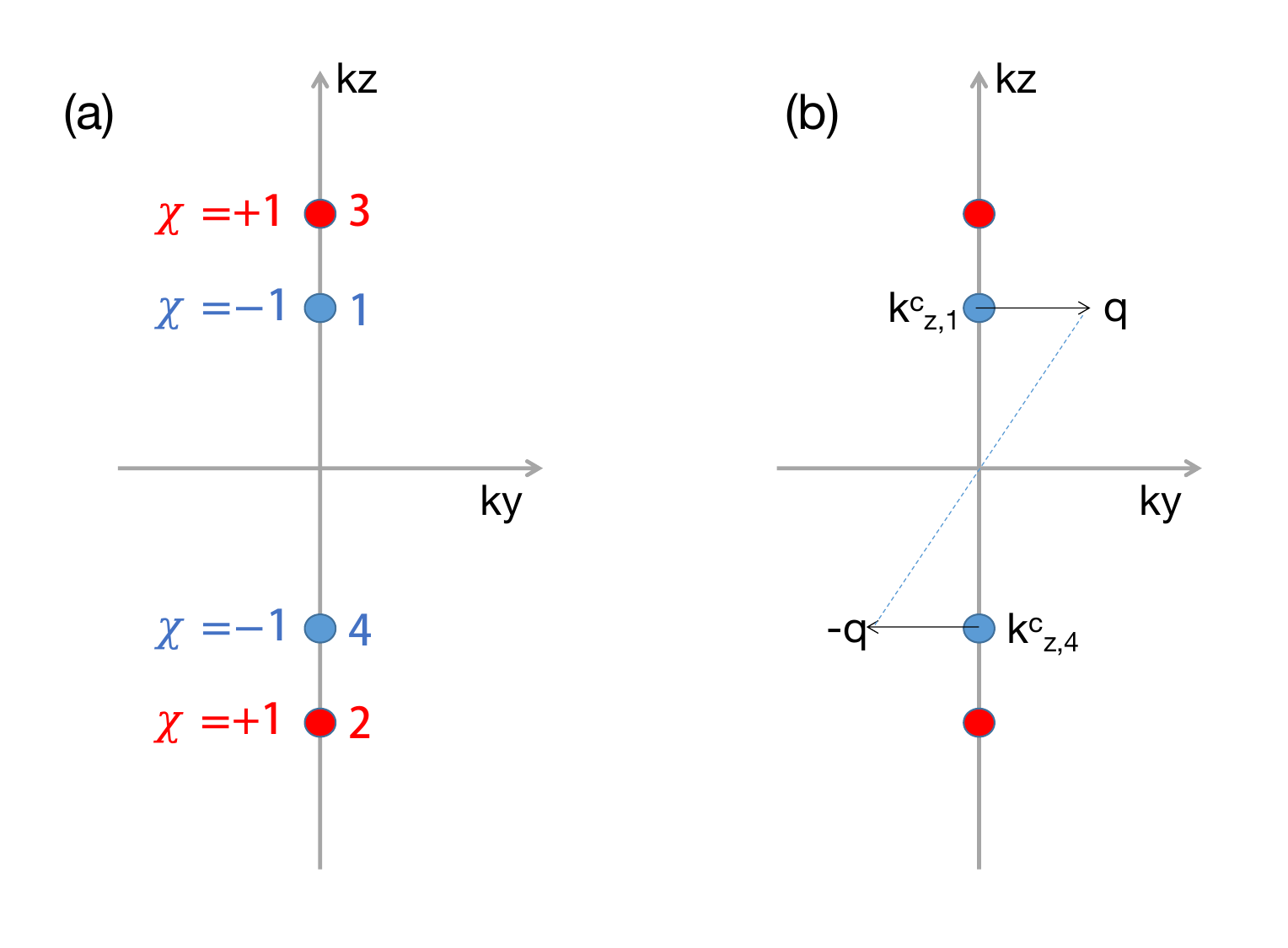}
	\caption{The distribution of Weyl points and the schematic figure of the BCS pairing. (a) The distribution of the four Weyl points in our linear Weyl semimetal. (b) The pairing between electrons around two Weyl points with same chirality and opposite momentum.}
	\label{AF2}
\end{figure}

In this section we will derive the effective BdG Hamiltonian  \cite{PhysRevB.86.214514,PhysRevB.86.054504} describing the low-energy physics of SC Weyl semimetals, based on which we will define the emergent second Chern number \cite{PhysRevB.82.115120,bernevig2013topological,PhysRevB.78.195424,PhysRevLett.118.207002} in a synthetic four-dimensional space. We first derive the effective BdG Hamiltonian of SC linear Weyl semimetal with vorticity $n=1$. Fig. \ref{AF2}(a) shows the distribution of the four Weyl points of linear Weyl semimetals.  We now focus on the effective BdG Hamiltonian with the pairing between electrons around Weyl points at $\textbf{k}^{c}_{1}=(0, 0, k^{c}_{z,1})$ and $\textbf{k}^{c}_{4}=(0, 0, k^{c}_{z,4})$, which is shown in Fig. \ref{AF2}(b).

The low-energy effective Hamiltonians around Weyl points $\textbf{k}^{c}_{1}$
(Eq. (\ref{C1})) and $\textbf{k}^{c}_{4}$ (Eq. (\ref{C2})) are derived in section S-3, based on which we can derive the effective BdG Hamiltonian. We consider the Nambu basis $\Psi_{\textbf{q}}=(c_{1,\textbf{q},+,\frac{1}{2}}, c_{1,\textbf{q},-,\frac{3}{2}},c^{\dagger}_{4,-\textbf{q},+,-\frac{1}{2}}, c^{\dagger}_{4,-\textbf{q},-,-\frac{3}{2}})^{T}$ and write the effective BdG Hamiltonians under this basis. Taking the pairing term into consideration, the BdG Hamiltonian can be written as $H_{\rm{BdG}}=\sum_{\textbf{q}} \Psi^{\dagger}_{\textbf{q}} H_{\rm{BdG}}(\textbf{q})\Psi_{\textbf{q}}$,
in which 
\begin{equation}
	\begin{aligned}
		H_{\rm{BdG}}(\textbf{q})=\left( \begin{matrix}  vq_{z} & Aq_{+}&\Delta&0\\  Aq_{-} & -vq_{z} &0&\Delta\\ \Delta&0&-vq_{z} & -Aq_{+}\\ 0&\Delta& -Aq_{-} & vq_{z}  \end{matrix}\right). 
		\label{C4}
	\end{aligned}
\end{equation}
Here  we have considered the BCS pairing term $H_{\rm{BCS}}=\sum_{\textbf{q}}  \Delta c^{\dagger}_{1,\textbf{q},+,\frac{1}{2}} c^{\dagger}_{4,-\textbf{q},+,-\frac{1}{2}}+ \Delta c^{\dagger}_{1,\textbf{q},-,\frac{3}{2}} c^{\dagger}_{4,-\textbf{q},-,-\frac{3}{2}}+ H.c.$, in which electrons with opposite momentum and angular momentum are paired together.
\par If we consider the $\pi$ flux inserted along $z$ direction, the effective BdG Hamiltonian will become 
\begin{equation}
	H_{\rm{BdG}}(\textbf{q})=\left( \begin{matrix}  vq_{z} & Aq_{+}&\Delta e^{i\phi}&0\\  Aq_{-} & -vq_{z} &0&\Delta e^{i\phi}\\ \Delta e^{-i\phi}&0&-vq_{z} & -Aq_{+}\\ 0&\Delta e^{-i\phi}& -Aq_{-} & vq_{z}  \end{matrix}\right),
	\label{C5}
\end{equation}
in which the phase $e^{i\phi}$ attached to $\Delta$ describes the effect of the vortex line. Similarly, we can obtain the effective BdG Hamiltonian for SC Weyl semimetals with chirality $\chi$ and vorticity $n$
\begin{equation}
	\begin{aligned}
		H_{\rm{BdG}}(\bm{q})&=\left( \begin{matrix}  vq_{z} & Aq^{\chi}_{-} &\Delta e^{in\phi}&0\\  Aq^{\chi}_{+} & -vq_{z} &0&\Delta e^{in\phi}\\ \Delta e^{-in\phi}&0&-vq_{z} & -Aq^{\chi}_{-} \\ 0&\Delta e^{-in\phi}& -Aq^{\chi}_{+} & vq_{z}  \end{matrix}\right),
		\label{E4}
	\end{aligned}
\end{equation}
in which $q_{i}$ $(i=x,y,z)$ is the momentum relative to a Weyl point, and $q_{\pm}=q_{x}\pm iq_{y}=qe^{\pm i\theta}$. 

If the Hamiltonian changes slowly as we change the parameter $\phi$, the three spatial variables  $q_{x}$, $q_{y}$, and $q_{z}$ together with the fourth variable $\phi$ constitute a synthetic four-dimensional space \cite{PhysRevB.82.115120,PhysRevLett.119.047001}, in which we can define the second Chern number $C_{2}$. To calculate $C_{2}$ of the above effective BdG Hamiltonian, we reformulate Eq. (\ref{E4}) as
\begin{equation}
	\begin{aligned}
		H_{\rm{BdG}}(\bm{q})&=Aq^{\chi}\cos(\chi\theta)\chi_{3}\otimes \tau_{1}+Aq^{\chi}\sin(\chi\theta)\chi_{3}\otimes \tau_{2} \\
		&+vq_{z}\chi_{3}\otimes \tau_{3} -\Delta \sin n\phi \chi_{2}\otimes \tau_{0} +\Delta \cos n\phi \chi_{1}\otimes \tau_{0}.
		\label{E45}
	\end{aligned}
\end{equation}
Then we define $d_{1}=Aq^{\chi}\cos(\chi\theta)$, $d_{2}=Aq^{\chi}\sin(\chi\theta)$, $d_{3}=vq_{z}$, $d_{4}=-\Delta \sin n\phi$, and $d_{5}=\Delta \cos n\phi$. In this way, we can write the effective BdG Hamiltonian as $H_{\rm{BdG}}=\sum_{i} d_{i}(\bm{q}, \phi)\cdot \Gamma_{i}$. Here $\Gamma_{i}$ are five mutually anti-commuting matrices in Eq. (\ref{E45}). In this case, the definition of the second Chern number is
\begin{equation}
	\begin{aligned}
		C_{2}&=\frac{3}{8\pi^{2}} \int d^{3}q \int d\phi  \frac{1}{d^{5}} \epsilon^{abcde} d_{a}\partial_{q}d_{b}\partial_{\theta}d_{c}\partial_{q_{z}}d_{d}\partial_{\phi}d_{e}.
		\label{E12}
	\end{aligned}
\end{equation}
Here we have defined  $d=(\sum_{i}d^{2}_{i})^{1/2}$.
If we substitute the explicit forms of the coefficients $d_{i}$ into Eq. (\ref{E12}), we will obtain
\begin{equation}
	\begin{aligned}
		C_{2}&=\frac{3}{8\pi^{2}} \int_{0}^{\infty} dq \int_{0}^{2\pi} d\theta \int^{\infty}_{-\infty}dq_{z} \int^{2\pi}_{0} d\phi  \frac{1}{d^{5}} \epsilon^{abcde} d_{a}\partial_{q}d_{b}\partial_{\theta}d_{c}\partial_{q_{z}}d_{d}\partial_{\phi}d_{e}\\
		&=\frac{3}{8\pi^{2}} \int_{0}^{\infty} dq \int_{0}^{2\pi} d\theta \int^{\infty}_{-\infty}dq_{z} \int^{2\pi}_{0} d\phi  \frac{1}{d^{5}}\cdot \{-A^{2}\chi^{2}v\Delta^{2}nq^{2\chi-1}\cos^{2}(n\phi) \cos^{2}(\chi\theta)\\
		&-A^{2}\chi^{2}v\Delta^{2}nq^{2\chi-1}\cos^{2}(n\phi) \sin^{2}(\chi\theta) -A^{2}\chi^{2}v\Delta^{2}nq^{2\chi-1}\sin^{2}(n\phi) \cos^{2}(\chi\theta)\\
		&-A^{2}\chi^{2}v\Delta^{2}nq^{2\chi-1}\sin^{2}(n\phi) \sin^{2}(\chi\theta)\}\\
	\end{aligned}
\end{equation}
The integrand can be written in  a more compact form
\begin{equation}
	\begin{aligned}
		C_{2}&=-\frac{3}{8\pi^{2}} A^{2} \chi^{2} v \Delta^{2} n  \int_{0}^{\infty} dq \int_{0}^{2\pi} d\theta \int^{\infty}_{-\infty}dq_{z} \int^{2\pi}_{0} d\phi  \frac{q^{2\chi-1}}{(A^{2}q^{2\chi}+v^{2}q^{2}_{z}+\Delta^{2})^{5/2}} \\
		&=-\frac{3}{8\pi^{2}} A^{2} \chi^{2} v  \Delta^{2} n  2\pi 2\pi   \int^{\infty}_{-\infty}dq_{z}  \int_{0}^{\infty} dq \frac{q^{2\chi-1}}{(A^{2}q^{2\chi}+v^{2}q^{2}_{z}+\Delta^{2})^{5/2}} \\
		&=-\frac{3}{8\pi^{2}}  \chi^{2} v  \Delta^{2} n  2\pi 2\pi \frac{1}{2\chi}  \int^{\infty}_{-\infty}dq_{z} \{-\frac{2}{3} \frac{1}{(x+v^{2}q^{2}_{z}+\Delta^{2})^{3/2}}|^{x=\infty}_{x=0}\}\\
		&=-\frac{3}{8\pi^{2}}  \chi^{2} v  \Delta^{2} n  2\pi 2\pi \frac{1}{2\chi}  \int^{\infty}_{-\infty}dq_{z} \frac{2}{3}\frac{1}{(v^{2}q^{2}_{z}+\Delta^{2})^{3/2}}\\
		&=-n\cdot \chi.
	\end{aligned}
\end{equation}
This is the expected topological invariant which describes the dependence of gapless vortex modes on the vorticity $n$ and chirality $\chi$ in SC linear and quadratic Weyl semimetals.

\section{Dependence of VBSs on Chirality $\chi$ and Vorticity $n$}
\begin{figure}[htbp]
	\centering
	\includegraphics[scale=0.5]{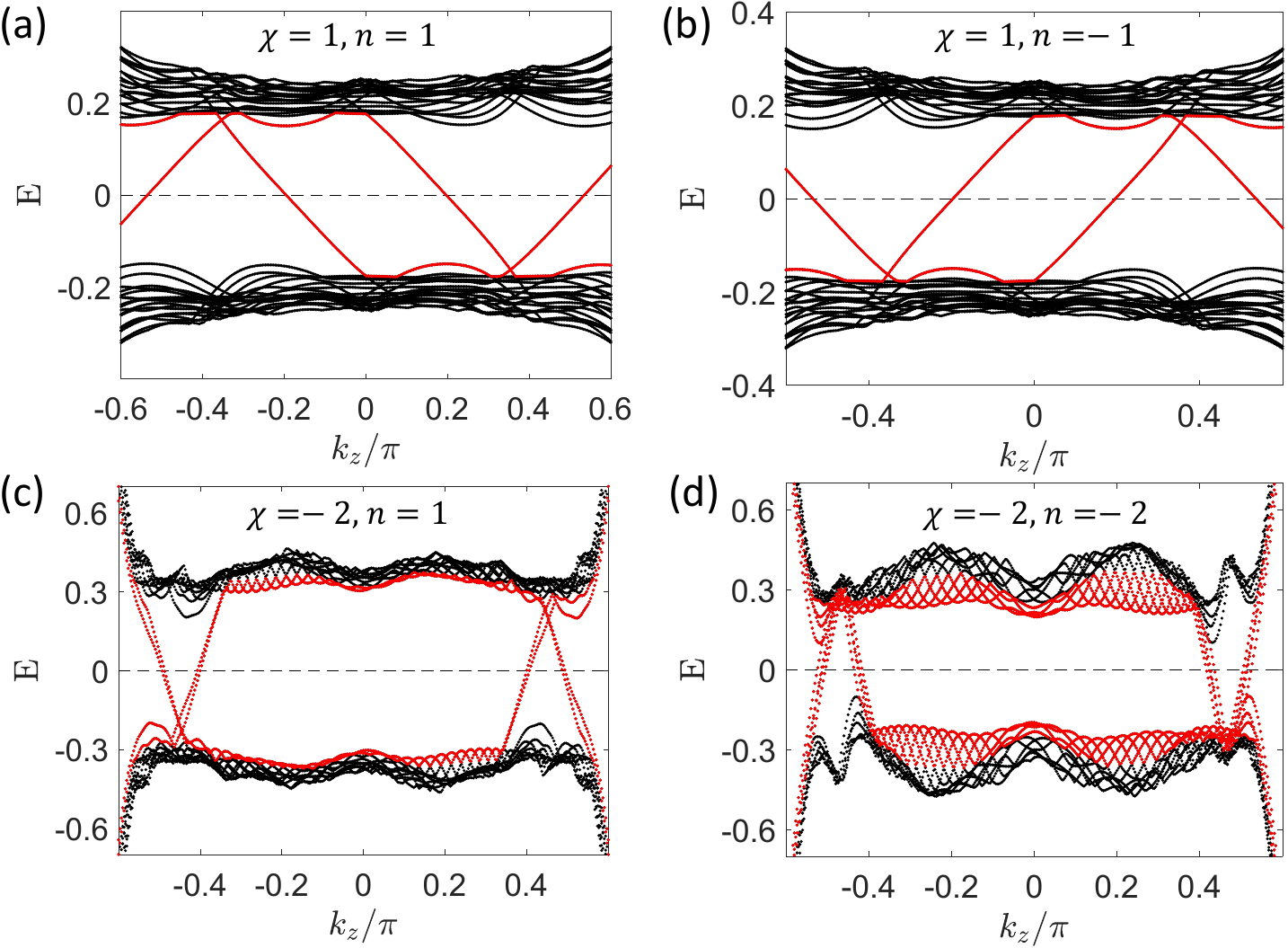}
	\caption{The VBSs of SC Weyl semimetals calculated with various chirality $\chi$ and vorticity $n$. (a) The VBS of SC linear Weyl semimetal  ($\chi=1$) are calculated with $\mu=0.2$, $\eta=0.3$, and $n=1$. (b)  The VBS of SC linear Weyl semimetal  ($\chi=1$) are calculated with $\mu=0.2$, $\eta=0.3$, and $n=-1$.   (c) The VBS of SC quadratic Weyl semimetal  ($\chi=-2$) are calculated with $\mu=0.4$, $\eta=0.7$, and $n=1$. (d) The VBS of SC quadratic Weyl semimetal  ($\chi=-2$) are calculated with $\mu=0.4$, $\eta=0.7$, and $n=-2$.}
	\label{F3}
\end{figure}

In this section we will discuss how the chirality $\chi$ and vorticity $n$ affect the numbers and propagating directions of gapless  modes in the vortex line of SC Weyl semimetals. Fig. \ref{F3}(a) and Fig. \ref{F3}(b) shows the VBSs ($E-k_{z}$ relation) of SC linear Weyl semimetal ($\chi=1$) with vorticity $n=1$ and $n=-1$, respectively. Fig. \ref{F3}(c) and Fig. \ref{F3}(d) shows the VBSs of SC quadratic Weyl semimetal ($\chi=-2$) with vorticity $n=1$ and $n=-2$, respectively. We find that there exist gapless  modes in each of the four sub-figures. 

By comparing the VBSs in two figures with same $n$ ($\chi$) and different $\chi$ ($n$), we can find how  $\chi$ ($n$) affects the propagating direction and number of gapless  modes around the $k_{z}$ component of one Weyl point.  It turns out that the number of gapless  modes around (the $k_{z}$ component of) each Weyl point is $\vert  n\chi \vert $, and the propagating direction of the gapless  modes around each Weyl point is determined by the sign of the product $-n\chi$. For example, there is one gapless  mode around each linear Weyl point in Fig. \ref{F3}(a), while there are two gapless  modes around each quadratic Weyl point in Fig. \ref{F3}(c), which verifies that the number of gapless  modes around each Weyl point is proportional to $\vert \chi\vert$. There are four gapless  modes around each quadratic Weyl point in Fig. \ref{F3}(d), while there are two gapless  modes around each Weyl point in Fig. \ref{F3}(c), which shows that the number of gapless  modes around each Weyl point is proportional to $\vert n\vert$.  The propagating directions of gapless  modes around each linear Weyl point in Fig. \ref{F3}(b) are opposite to those in Fig. \ref{F3}(a), showing that the propagating directions of gapless  modes are proportional to the sign of $n$. We know that in our models the chirality of two inner Weyl points is opposite to that of two outer Weyl points on the $k_{z}$ axis. As a result, in each of the four sub-figures in Fig. \ref{F3} the propagating direction of the inner two branches of gapless  modes is opposite to that of the outer  two branches of gapless  modes, which shows that the propagating directions of the gapless  modes are proportional to the sign of $\chi$. 

\section{The Tight-binding Models of Linear and Quadratic Weyl Semimetals}


In the main text, we have given the $\textbf{k}\cdot \textbf{p}$ model of linear and  quadratic Weyl semimetals. However, we use the lattice models rather than the $\bm{k}\cdot \bm{p}$ models to calculate the vortex bound states. So in this section, we introduce the tight-binding lattice models  of linear and quadratic Weyl semimetals. Since Dirac semimetals are well classified \cite{BJYang}, we decide to break inversion symmetry in a Dirac semimetal to obtain a Weyl semimetal.
\subsection{The Lattice Model and $\textbf{k}\cdot \textbf{p}$ Model of a Linear Weyl Semimetal}
The tight-binding model of the linear Dirac semimetal  with TRS, inversion symmetry, and  $C^{z}_{4}$ rotational symmetry is written under the basis $(\vert+,\frac{1}{2} \rangle, \vert-,\frac{3}{2} \rangle, \vert+,-\frac{1}{2} \rangle, \vert-,-\frac{3}{2} \rangle)^{T}$, in which each state is the eigenstate of inversion and $C^{z}_{4}$ operations labelled by the corresponding eigenvalues. Adding the inversion symmetry breaking term $\eta \sin k_{z}\sigma_{3}\otimes \tau_{3}$ into the model of linear Dirac semimetal leads to the tigh-binding lattice model of linear Weyl semimetal 
\begin{equation}
\begin{aligned}
&H_{\rm{LWSM}}(\textbf{k})\\
&=(m+2t^{A}_{1}\cos k_{x}+2t^{A}_{1}\cos k_{y}+2t^{D}_{1}\cos k_{z})\sigma_{0}\otimes \tau_{3}\\
&+2t^{A}_{3}\sin k_{x} \sigma_{3}\otimes \tau_{1}-2t^{A}_{3}\sin k_{y} \sigma_{0}\otimes \tau_{2}\\
&-8t^{C}_{4}\sin k_{z}(\cos k_{x}-\cos k_{y}) \sigma_{1}\otimes \tau_{1}\\
&+8t^{B}_{4}\sin k_{x}\sin k_{y} \sin k_{z}\sigma_{2}\otimes \tau_{1}+\eta \sin k_{z}\sigma_{3}\otimes \tau_{3}.
\label{A1}
\end{aligned}
\end{equation}
Here we consider the cubic lattice. $t^{A}_{1}$ and $t^{A}_{3}$ are the nearest-neighbor hopping terms in $x-y$ plane, and $t^{D}_{1}$ is the nearest neighbor hopping term in $z$ direction. $t^{C}_{4}$ is the next nearest neighbor hopping term, and $t^{B}_{4}$ is the third nearest neighbor hopping term. $m$ and $\eta$ are the on-site term and inversion symmetry breaking term, respectively. Here  $\sigma$ and $\tau$ represent spin and orbital degrees of freedom, respectively. 

If we substitute $\sin k_{i}$ with $k_{i}$ and substitute $\cos k_{i}$ with $1-\frac{1}{2}k^{2}_{i}$ ($i=x$, $ y$, and $z$), we will obtain the  $\textbf{k}\cdot \textbf{p}$ model of linear Weyl semimetal
\begin{equation}
	\begin{aligned}
		H_{\rm{LWSM}}(\bm{k})=\left( \begin{matrix} M(\bm{k})+\eta k_{z} & Ak_{+}&0 & B^{*}(\bm{k})\\  Ak_{-}& -M(\bm{k})-\eta k_{z}& B^{*}(\bm{k}) &0 \\ 0 &B(\bm{k}) &M(\bm{k})-\eta k_{z} &-Ak_{-}\\  B(\bm{k}) &0 &-Ak_{+} &-M(\bm{k})+\eta k_{z}  \end{matrix} \right)
	\end{aligned}.
	\label{E1}
\end{equation}
Here  $M(\bm{k})=M_{0}-M_{1}k^{2}_{z}-M_{2}(k^{2}_{x}+k^{2}_{y})$, and $B(\bm{k})=\beta k_{z}(k^{2}_{x}-k^{2}_{y})+2i\alpha k_{x}k_{y}k_{z}$. The correspondence between parameters in the $\textbf{k}\cdot \textbf{p}$ model and  those in the lattice model is $A=2t^{A}_{3}$, $\alpha=4t^{B}_{4}$, $\beta=4t^{C}_{4}$, $M_{0}=m+4t^{A}_{1}+2t^{D}_{1}$, $M_{1}=t^{D}_{1}$, and $M_{2}=t^{A}_{1}$. 

Next we show that the low-energy effective Hamiltonian is a Weyl Hamiltonian. The gap closing condition of above Hamiltonian leads to four Weyl points located on $k_{z}$ axis with $k^{c}_{z1,2}=(-\eta\pm \sqrt{\eta^{2}+4M_{0}M_{1}})/(-2M_{1})$ and $k^{c}_{z3,4}=(\eta\pm \sqrt{\eta^{2}+4M_{0}M_{1}})/(-2M_{1})$. When we consider the low-energy physics, we omit the higher-order term $B(\textbf{k})$ in the $\textbf{k}\cdot \textbf{p}$ model of linear Weyl semimetal. Then the Hamiltonian will be decoupled into the $2\times 2$ upper-left block and the $2\times2$ lower-right block. The Weyl point at $\textbf{k}^{c}_{1}=(0, 0, k^{c}_{z1})$ is obtained from the gap closing condition in the upper-left block. So we expand the upper-left block of the Hamiltonian around the Weyl point located at $\textbf{k}^{c}_{1}$
\begin{equation}
	\begin{aligned}
		&H_{\rm{UL}}(\textbf{k}^{c}_{1}+\delta \textbf{k})\\
		&=\left( \begin{matrix} (\eta-2M_{1}k^{c}_{z,1})\delta k_{z} & A(\delta k_{x}+i\delta k_{y}) \\  A(\delta k_{x}-i\delta k_{y}) & -(\eta-2M_{1}k^{c}_{z,1})\delta k_{z}  \end{matrix}\right)\\
		&=\left( \begin{matrix} (\eta-2M_{1}k^{c}_{z,1})q_{z} & A(q_{x}+iq_{y}) \\  A(q_{x}-iq_{y}) & -(\eta-2M_{1}k^{c}_{z,1})q_{z}  \end{matrix}\right)\\
		&=\left( \begin{matrix} vq_{z} & A(q_{x}+iq_{y}) \\  A(q_{x}-iq_{y}) & -vq_{z}  \end{matrix}\right).
		\label{C1}
	\end{aligned}
\end{equation}
Here we use the notation $\textbf{q}=\delta \textbf{k}$ and $v=\eta-2M_{1}k^{c}_{z,1}=\sqrt{\eta^{2}+4M_{0}M_{1}}$. The Weyl point at $\bm{k^{c}_{4}}=(0, 0, k^{c}_{z4})$ is obtained from the gap closing condition in the lower-right block.
Then we expand the lower-right block around $\textbf{k}^{c}_{4}$
\begin{equation}
	\begin{aligned}
		&H_{\rm{LR}}(\textbf{k}^{c}_{4}+\delta \textbf{k})\\
		&=\left( \begin{matrix} (-\eta-2M_{1}k^{c}_{z,4})\delta k_{z} & -A(\delta k_{x}-i\delta k_{y}) \\  -A(\delta k_{x}+i\delta k_{y}) & (\eta+2M_{1}k^{c}_{z,4})\delta k_{z}  \end{matrix}\right)\\
		&=\left( \begin{matrix} (-\eta-2M_{1}k^{c}_{z,4})q_{z} & -A(q_{x}-iq_{y}) \\  -A(q_{x}+iq_{y}) & (\eta+2M_{1}k^{c}_{z,4})q_{z}  \end{matrix}\right)\\
		&=\left( \begin{matrix} -vq_{z} & -A(q_{x}-iq_{y}) \\  -A(q_{x}+iq_{y}) & vq_{z}  \end{matrix}\right).
		\label{C2}
	\end{aligned}
\end{equation}
Here we use the notation $\textbf{q}=\delta \textbf{k}$ and $v=\eta+2M_{1}k^{c}_{z,4}=\sqrt{\eta^{2}+4M_{0}M_{1}}$. Obviously, Eq. (\ref{C1}) and Eq. (\ref{C2}) describe linear Weyl fermions. In fact, the low-energy Hamiltonians around other two Weyl points $\textbf{k}^{c}_{2}$ and $\textbf{k}^{c}_{3}$ are also Weyl Hamiltonians. Thus, Eq. (\ref{E1}) is indeed a Hamiltonian describing a Weyl semimetal.

\subsection{The Lattice Model and $\textbf{k}\cdot \textbf{p}$ Model of Quadratic Weyl Semimetal}
%
According to the classification of Dirac semimetals \cite{BJYang}, the appearance of a quadratic Dirac point requires the existence of $C^{z}_{6}$ symmetry.  Therefore, the  lattice model of the quadratic Dirac semimetal is written under the basis $(\vert+,\frac{5}{2} \rangle, \vert-,\frac{1}{2} \rangle, \vert+,-\frac{5}{2} \rangle, \vert-,-\frac{1}{2} \rangle)^{T}$, in which each state is the eigenstate of inversion and $C^{z}_{6}$ operations labelled by the corresponding eigenvalues. After considering an inversion symmetry breaking term $\eta\sin k_{z}\sigma_{3}\otimes \tau_{3}$, we write down the tight-binding lattice model of a quadratic Weyl semimetal
\begin{equation}
\begin{aligned}
&H_{\rm{QWSM}}(\textbf{k})\\
&=\{m+2t^{A}_{1}[\cos k_{x}+2\cos \frac{1}{2}k_{x} \cos \frac{\sqrt{3}}{2}k_{y}]+2t^{D}_{1}\cos k_{z} \}\cdot \\
&\sigma_{0}\otimes \tau_{3} -2t^{A}_{4}[\sin k_{x}-2\sin \frac{1}{2}k_{x}\cos \frac{\sqrt{3}}{2}k_{y}]\sigma_{1}\otimes \tau_{1}\\
&-2t^{B}_{4}(2\cos \frac{3}{2}k_{x}\sin \frac{\sqrt{3}}{2}k_{y}-\sin \sqrt{3}k_{y}) \sigma_{2}\otimes \tau_{1}\\
&-4t^{C}_{3}\sin k_{z} (\cos k_{x}
-\cos \frac{1}{2}k_{x}\cos \frac{\sqrt{3}}{2}k_{y})\sigma_{3}\otimes \tau_{1}\\
&+4\sqrt{3}t^{C}_{3}\sin k_{z}\sin \frac{1}{2} k_{x}\sin \frac{\sqrt{3}}{2}k_{y} \sigma_{0}\otimes \tau_{2}\\
&+\eta\sin k_{z}\sigma_{3}\otimes \tau_{3}.
\label{A2}
\end{aligned}
\end{equation}
Here we consider the three-dimensional hexagonal lattice. $t^{A}_{1}$ and $t^{A}_{4}$ are the nearest-neighbor hopping terms in $x-y$ plane, and $t^{D}_{1}$ is the nearest neighbor hopping term in $z$ direction. $t^{B}_{4}$ is the third nearest neighbor hopping term in $x-y$ plane, and $t^{C}_{3}$ is the next nearest neighbor hopping term. $m$ and $\eta$ are the on-site term and inversion symmetry-breaking term, respectively.  Here  $\sigma$ and $\tau$ represent spin and orbital degrees of freedom, respectively.  

If we substitute $\sin k_{i}$ with $k_{i}$ and substitute $\cos k_{i}$ with $1-\frac{1}{2}k^{2}_{i}$ ($i=x$, $ y$, and $z$), we will obtain the  $\textbf{k}\cdot \textbf{p}$ model of quadratic Weyl semimetal
\begin{equation}
	\begin{aligned}
		H_{\rm{QWSM}}(\bm{k})=\left( \begin{matrix} M(\bm{k})+\eta k_{z} & Ak^{2}_{-}k_{z} &0 & B^{*}(\bm{k})\\  Ak^{2}_{+}k_{z}& -M(\bm{k})-\eta k_{z}& B^{*}(\bm{k}) &0 \\ 0 &B(\bm{k}) &M(\bm{k})-\eta k_{z} &-Ak^{2}_{+}k_{z}\\  B(\bm{k}) &0 &-Ak^{2}_{-}k_{z} &-M(\bm{k})+\eta k_{z}  \end{matrix} \right)
	\end{aligned}.
	\label{E5}
\end{equation}
Here  $M(\bm{k})=M_{0}-M_{1}k^{2}_{z}-M_{2}(k^{2}_{x}+k^{2}_{y})$, and $B(\bm{k})=\alpha k_{x}(k^{2}_{x}-3k^{2}_{y})+i\beta k_{y}(3k^{2}_{x}-k^{2}_{y})$. The correspondence between parameters in the $\textbf{k}\cdot \textbf{p}$ model and  those in the lattice model is $A=\frac{3}{2}t^{C}_{3}$, $\alpha=\frac{1}{4}t^{A}_{4}$, $\beta=2t^{B}_{4}(\frac{\sqrt{3}}{2})^{3}$,  $M_{0}=m+6t^{A}_{1}+2t^{D}_{1}$, $M_{1}=t^{D}_{1}$, and $M_{2}=\frac{3}{2}t^{A}_{1}$. 

Next we show that the corresponding low-energy effective Hamiltonian is a quadratic Weyl Hamiltonian. The gap closing condition of above Hamiltonian leads to four Weyl points located on $k_{z}$ axis with $k^{c}_{z1,2}=(-\eta\pm \sqrt{\eta^{2}+4M_{0}M_{1}})/(-2M_{1})$ and $k^{c}_{z3,4}=(\eta\pm \sqrt{\eta^{2}+4M_{0}M_{1}})/(-2M_{1})$. When we consider the low-energy physics, we omit the higher-order term $B(\textbf{k})$ in the $\textbf{k}\cdot \textbf{p}$ model of quadratic Weyl semimetal. Then the Hamiltonian will be decoupled into the $2\times 2$ upper-left block and the $2\times2$ lower-right block. The Weyl point at $\textbf{k}^{c}_{1}=(0, 0, k^{c}_{z1})$ is obtained from the gap closing condition in the upper-left block. So we expand the upper-left block of the Hamiltonian around the Weyl point located at $\textbf{k}^{c}_{1}$
\begin{equation}
	\begin{aligned}
		&H_{\rm{UL}}(\textbf{k}^{c}_{1}+\delta \textbf{k})\\
		&=\left( \begin{matrix} (\eta-2M_{1}k^{c}_{z,1})\delta k_{z} & Ak^{c}_{z,1}(\delta k_{x}-i\delta k_{y})^2 \\  Ak^{c}_{z,1}(\delta k_{x}+i\delta k_{y})^{2} & -(\eta-2M_{1}k^{c}_{z,1})\delta k_{z}  \end{matrix}\right)\\
		&=\left( \begin{matrix} (\eta-2M_{1}k^{c}_{z,1})q_{z} & Ak^{c}_{z,1}(q_{x}-iq_{y})^{2} \\  Ak^{c}_{z,1}(q_{x}+iq_{y})^{2} & -(\eta-2M_{1}k^{c}_{z,1})q_{z}  \end{matrix}\right)\\
		&=\left( \begin{matrix} vq_{z} & Ak^{c}_{z,1}(q_{x}-iq_{y})^{2} \\  Ak^{c}_{z,1}(q_{x}+iq_{y})^{2} & -vq_{z}  \end{matrix}\right).
		\label{C3}
	\end{aligned}
\end{equation}
Here we use the notation $\textbf{q}=\delta \textbf{k}$ and $v=\eta-2M_{1}k^{c}_{z,1}=\sqrt{\eta^{2}+4M_{0}M_{1}}$. We find that Eq. (\ref{C3}) describes quadratic Weyl fermions. It is not hard to derive the low-energy Hamiltonians around other three Weyl points, and all of them are quadratic Weyl Hamiltonians. Therefore, the Hamiltonian Eq. (\ref{E5}) is the model describing a quadratic Weyl semimetal.

\bibliography{supp_refs}
\bibliographystyle{apsrev4-2}